\documentclass[lettersize,journal]{IEEEtran}
\usepackage{amsmath,amsfonts}
\usepackage{bm}
\usepackage{algorithmic}
\usepackage{algorithm}
\usepackage{array}
\usepackage{textcomp}
\usepackage{stfloats}
\usepackage{url}
\usepackage{verbatim}
\usepackage{graphicx}
\usepackage{cite}
\usepackage{enumitem}

\begin{document}

\title{Peer-to-Peer Cloud Service Market for Data Centers Oriented to Computation-Electricity Coordination}
\author{Yugui Liu, Yibo Ding,~\IEEEmembership{Graduate Student Member,~IEEE,}  Xudong Li, Jing Qu, 

Wenyi Zhang,~\IEEEmembership{Member,~IEEE,} Tong Qian,~\IEEEmembership{Member,~IEEE,} Wuyou Xiao,

Zhengyang Hu,~\IEEEmembership{Member,~IEEE,}  Jiaqi Ruan,~\IEEEmembership{Member,~IEEE,} Zhao Xu,~\IEEEmembership{Senior Member,~IEEE}}

\markboth{Journal of \LaTeX\ Class Files,~Vol.~14, No.~8, August~2021}%
{Shell \MakeLowercase{\textit{et al.}}: A Sample Article Using IEEEtran.cls for IEEE Journals}


\maketitle

\begin{abstract}
Energy-intensive data centers (DCs) have emerged as substantial and flexible loads in modern power systems, underscoring the critical need for computation-electricity coordination. Harnessing the spatio-temporal flexibility of DC workloads is a promising approach to facilitate this coordination. However, existing studies overlook the collaborative potential of computational resource sharing among geo-distributed DCs, thereby failing to fully unlock this flexibility.
In this paper, a bi-level computation-electricity coordination framework is proposed to explicitly capture the bidirectional interactions between DCs and power grid.
Firstly, a peer-to-peer cloud service market (P2P-CSM) for geo-distributed DCs is proposed, which enables bilateral cloud service transactions to leverage regional heterogeneities (e.g., electricity prices, cooling efficiency).
Secondly, locational marginal prices are embedded into the framework to reflect network congestion and nodal price disparities.
Thirdly, a dual consensus alternating direction method of multipliers (ADMM)-based decentralized algorithm is developed as the P2P market clearing algorithm, and a bisection-assisted iterative algorithm is proposed to ensure rigorous convergence of the framework.
Case studies conducted on modified IEEE 30-bus system validate that the P2P-CSM achieves a win-win computation-electricity coordination: it not only increases total DC operational profit by 22.8\%, but also effectively alleviates grid congestion and yields a 3.2\% reduction in total energy consumption.

\end{abstract}

\begin{IEEEkeywords}
Data center, computation-electricity coordination, cloud service, workload management, P2P, ADMM
\end{IEEEkeywords}

\section{Introduction}
The proliferation of artificial intelligence has catalyzed an exponential expansion of data centers (DCs).
According to the International Energy Agency, DCs accounted for approximately 415 TWh of global electricity consumption in 2024, nearly 1.5\% of global electricity demand, and this figure is projected to double to 3\% by 2030 \cite{EnergyAIAnalysis2025}. 
As energy-intensive entities, DCs increasingly play a pivotal role in modern power systems, influencing the secure and economic operation of the grid, thereby underscoring the critical need for computation-electricity coordination.

DC energy management serves as a fundamental technical pathway to operationalize such coordination. Existing research on DC energy management can be broadly classified into three principal domains.
The first category concentrates on the coordination of local distributed energy resources (DERs), such as photovoltaics (PV), wind power, and battery energy storage systems (BESS) \cite{kumarRenewableEnergybasedMultiindexed2018, zhouOptimalEnergyManagement2023,sunBatteryAssistedOnlineOperation2023, liEquilibriumbasedElectricitycomputationCollaborative2026}.
and the second addresses thermal management, encompassing air conditioning system inertia, thermal storage, etc. \cite{dayarathnaDataCenterEnergy2015, niReviewAirConditioning2017, dingIntegratedStochasticEnergy2019a, zhengTEShaveReducingData2015,ruanPrivacyPreservingBiLevelOptimization2024a}.
While these two domains are effective in reducing operational costs, the third and the most pivotal dimension is workload management, given that the IT equipment responsible for workload execution accounts for approximately 60\% of total electricity consumption \cite{dayarathnaDataCenterEnergy2015}.

DC workloads are typically categorized into batch and interactive types, each exhibiting distinct temporal and spatial characteristics \cite{liDecentralizedOptimizationIntegrated2023b}.
Batch workloads (BWs) (e.g., data analytics, model training ) are delay-tolerant and can be deferred over time scales from minutes to days. \cite{liuOptimalEnergyManagement2024} established a time-shifting model for BWs to jointly minimize operational costs and renewable curtailment. \cite{dingDataCenterJob2025} regulated power consumption by delaying BWs execution within Quality of Service (QoS). \cite{wangCoordinatedOptimalScheduling2023} formulated a BWs scheduling model that preserves time windows to account for their arrival times, execution times, and deadlines.
However, most existing workload shifting strategies primarily exploit the temporal flexibility of BWs, failing to leverage the spatial flexibility inherent in geo-distributed DC networks.

Underpinned by advancements in optical fiber communication and low-latency network architectures, interactive workloads (IWs) (e.g., web requests, online inference) could be migrated across geo-distributed DCs while satisfying QoS requirements \cite{kumarRenewableEnergyBasedMultiIndexed2019b}. 
This spatial flexibility unlocks the economic potential of underutilized computing resources, essentially transforming computing capacity into a spatially dispatchable resource, referred to as ‘cloud service’.
From a market perspective, due to its inherent divisibility and measurability, cloud service resource could be viewed as a tradable commodity analogous to electricity or energy capacity \cite{schweppeSpotPricingElectricity2013}.
Building on cloud federation theory \cite{darzanosCloudFederationsEconomics2019}, workload migration enables DCs to function as ‘Prosumers’ (both cloud service suppliers and consumers) to maximize economic benefits within a collaborative federation.
Consequently, the operational paradigm of DCs is redefined, and inter-DC workload migration is no longer merely a technical operation but a cloud service commodity transaction in a market. 


However, a challenge remains in how geo-distributed DCs, which maybe operated by different entities, can effectively coordinate cloud service trading in practice.  
Each operator inherently seeks its decision-making autonomy and privacy confidentiality.
A peer-to-peer (P2P) market offers a promising decentralized paradigm, enabling DCs to negotiate bilaterally and execute transactions without relying on a central coordinator.
To this end, we propose a P2P-Cloud Service Market (P2P-CSM) for geo-distributed DCs as a practical coordination mechanism. Within this market, a DC can purchase cloud services to migrate its IWs or share surplus capacity with peers, thereby capitalizing on regional heterogeneity (e.g., electricity prices,  cooling efficiency and renewable availability) to enhance overall operational benefits.

To address P2P trading problems, several distributed methods have been explored, including the dual ascent method \cite{fengPeertoPeerEnergyTrading2023}, the generalized Benders decomposition \cite{xiaPreservingOperationPrivacy2022}, and the Alternating Direction Method of Multipliers (ADMM) \cite{ullahPeertoPeerEnergyTrading2021}, etc. Notably, the ADMM-based distributed algorithm has been extensively adopted due to stable convergence properties and broad applicability. However, standard ADMM implementations still require a central coordinator and fail to achieve full decentralization. 

In addition, since electricity procurement constitutes the dominant portion of DC operational costs, electricity price signals are widely used to guide computation-electricity coordination. 
\cite{hogadeMinimizingEnergyCosts2018} optimized workload allocation based on time-of-use pricing to minimize electricity costs. \cite{chenProliferationSmallData2022a} introduced an aggregation-based approach that exploits spatial diversity in regional electricity prices. These works treat DCs as price takers, responding to exogenous pricing signals; however, the massive electricity demand of DCs may possess market power to influence electricity prices, which has motivated the development of price-maker models. \cite{wangProactiveDemandResponse2016} investigated how workload redistribution alters grid load ratios and clearing prices through a two-stage optimization model. \cite{tranHowGeoDistributedData2016} formulated a Stackelberg game to model the strategic response of DCs to grid pricing through workload scheduling. But these studies rely on simplified electricity price models that neglect network constraints and nodal price disparities.

However, geo-distributed DCs within the same region typically participate in a unified electricity market and are subject to locational marginal prices (LMPs), which are widely implemented in the U.S. \cite{litvinovElectricityMarketsUnited2019}. 
By determining node-specific prices based on supply-demand balance and network congestion, LMP serves as a vital signal for computation-electricity coordination: it guides DC workload scheduling, which in turn reshapes the LMPs. Explicitly modeling this bidirectional coordination could not only maximize the economic profitability of DCs but also serve as a proactive tool to alleviate transmission congestion, thereby safeguarding the secure and stable operation of the power system.

In consideration of the aforementioned issues, the main contributions of this paper are summarized as follows:
\begin{enumerate}[label=(\arabic*), leftmargin=2em]
     \item \textbf{Decentralized cloud service trading mechanism}: A novel P2P-CSM is proposed to enable bilateral cloud service transactions among geo-distributed DCs without a central coordinator. Under this mechanism, DCs can exploit regional heterogeneities (e.g., electricity prices, cooling efficiency) through spatial workload migration,  thereby maximizing their operational profitability.
    \item \textbf{LMP-embedded computation-electricity coordination framework}: A bi-level coordination framework is proposed: the upper level solves optimal power flow to determine LMPs, and the lower level optimizes DC workload scheduling and P2P-CSM trading strategies accordingly, which, in turn, updates the nodal prices. LMPs are embedded into the framework to reflect network congestion and nodal disparities. 
    \item \textbf{Solving algorithms}: A dual consensus ADMM-based decentralized algorithm is developed as the P2P market clearing algorithm, and a bisection-assisted iterative algorithm is proposed to ensure rigorous convergence of the computation-electricity coordination framework.
\end{enumerate}

\begin{figure}[t]
    \centering
    \includegraphics[width=0.9\linewidth]{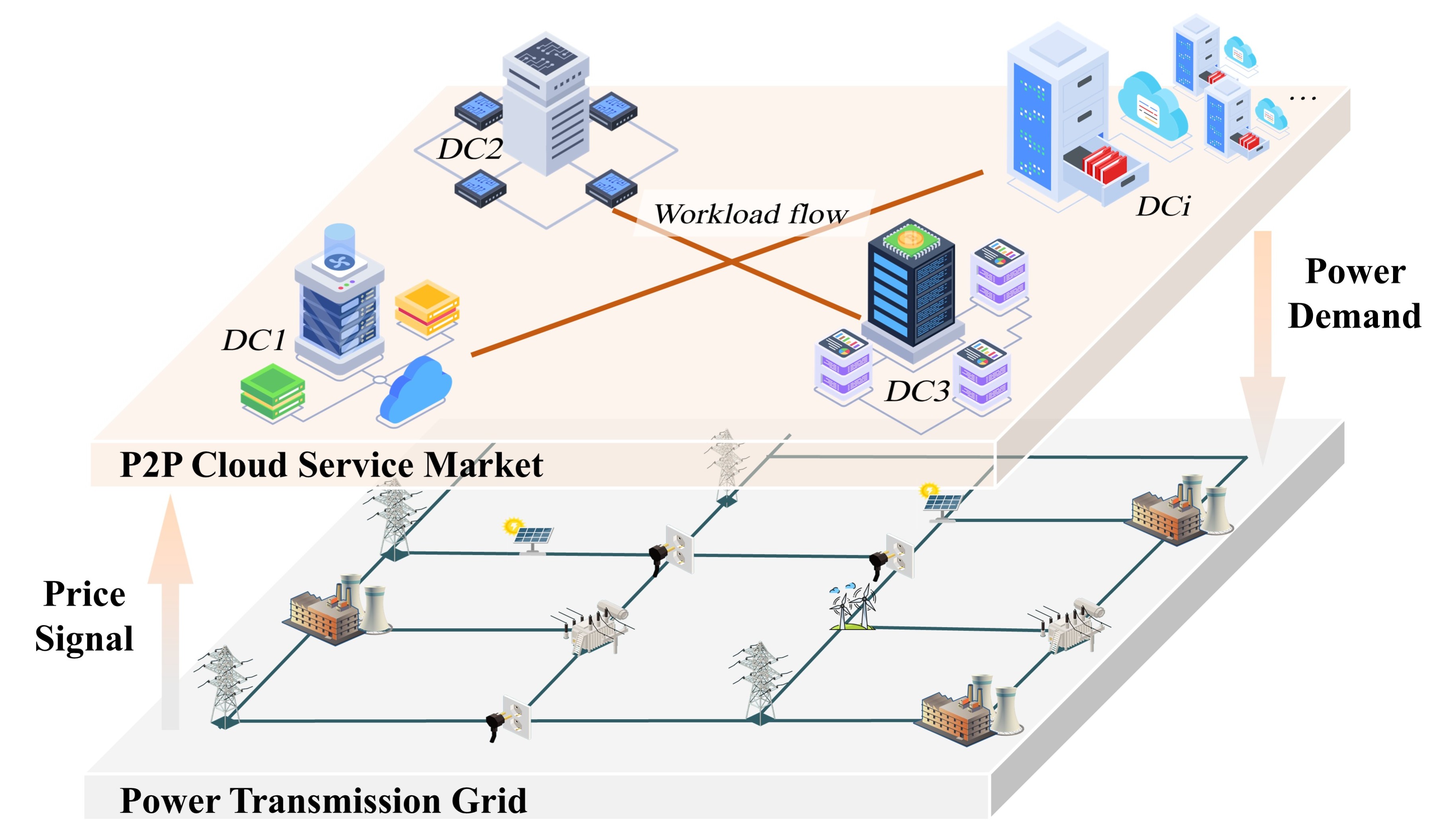}   
    \caption{The proposed electricity–computation coordination framework}
    \label{fig:placeholder}
\end{figure}

This paper is organized as follows. Section \ref{Problem Description} is the problem description. Section \ref{Mathematical Formulation} presents the detailed mathematical formulations of the lower and upper level models. The proposed P2P-CSM and the corresponding problem reformulation are introduced in Section \ref{Problem Reformulation with P2P-CSM}. Section \ref{solution} presents the solution methodology. Section \ref{Case Studies} conducts case studies and results analysis. Finally, Section \ref{Conclusion} concludes the paper.

\section{Problem Description}
\label{Problem Description}
The proposed bi-level computation-electricity coordination framework is illustrated in Fig.\ref{fig:placeholder}. 
We consider a set of geo-distributed DCs, each located at a different bus of the power grid and interconnected via high-speed optical fiber communication networks.
Let $\mathcal{I} = \{1,2,\ldots,I\}$ denote the set of DCs, and $\mathcal{I}_i$ denote the neighborhood of DC~$i$.
In this paper, the coordination is formulated at the day-ahead scheduling stage. The power grid operator provides LMPs to DCs. Guided by these nodal prices, each DC optimizes its workload scheduling and P2P trading strategy. The resulting DC power demand is then fed back to the power grid operator for LMP recalculation. Through this LMP-feedback process, the proposed framework captures the mutual interaction between profit-oriented DC operation and secure and economic power system operation.


\section{Mathematical Formulation}
\label{Mathematical Formulation}
In this section, the detailed mathematical formulations, including the DC energy and workload management model (lower level) and the power grid pricing model (upper level), will be presented.
\subsection{Data Center Energy and Workload Management}
The power consumption of each DC can be decomposed into three major components: the IT system, the cooling system, and other auxiliary facilities \cite{chenInternetDataCenter2021}. Accordingly, the total power consumption of DC $i$ at time $t$  is expressed as \eqref{eq:dc_total_power}.
\begin{equation}
P_{i,t}^{\mathrm{DC}} = P_{i,t}^{\mathrm{IT}} + P_{i,t}^{\mathrm{C}} + P_{i,t}^{\mathrm{Aux}}
\label{eq:dc_total_power}
\end{equation}
where $P_{i,t}^{\mathrm{IT}}$ denotes the IT system power consumption, $P_{i,t}^{\mathrm{C}}$ denotes the cooling power consumption and $P_{i,t}^{\mathrm{Aux}}$ denotes to the auxiliary facility power consumption.
\subsubsection{ IT System Power Consumption Model}
To more accurately characterize the IT system's power consumption, this paper will not only consider servers' power consumption but also that of the underlying network communication infrastructure.
\begin{equation}
P_{i,t}^{\mathrm{IT}} = 
\left(e_{i}^{\mathrm{N}} + e_i^{I}\right)m_{i,t}
+ \frac{e_i^{P} - e_i^{I}}{\gamma_i} w_{i,t}
\label{eq:it_power}
\end{equation}  
\begin{equation}
e_{i}^{\mathrm{N}} = 
\frac{n_i^{E} e_i^{E} + n_i^{A} e_i^{A} + n_i^{C} e_i^{C}}{M_i}
\label{eq:en}
\end{equation}  
where \(m_{i,t}\) denotes the number of active servers of DC $i$ at time $t$, \(w_{i,t}\) is the processed workload of DC $i$ at time $t$, \(e_i^{I}\) and \(e_i^{P}\) represent the idle and peak power of server in DC $i$, and \(\gamma_i\) denotes the average service rate. 
Especially, \(e_i^{\mathrm{N}}\) denotes the network communication power between network equipments \eqref{eq:en}, where \(n_i^{E}\), \(n_i^{A}\), \(n_i^{C}\) denote the number of edge switches, aggregation switches, and core switches in DC \(i\), respectively; \(e_i^{E}, e_i^{A}, e_i^{C}\) are the corresponding power consumption, \(M_i\) is the total number of servers in DC \(i\).  

\subsubsection{Cooling System Power Consumption Model}
To maintain appropriate operating conditions, DCs rely on cooling systems to dissipate the heat generated by IT equipment and auxiliary devices. Unlike simplified models that use a fixed power usage effectiveness (PUE), we explicitly model cooling power consumption based on cooling facility's characteristics and ambient conditions, shown as \eqref{eq:cool_power}-\eqref{eq:indoor_temp_bounds}.

\begin{equation}
P_{i,t}^{\mathrm{C}} = \frac{H_{i,t}}{\beta_i} + P_{i}^{\mathrm{Cmin}}
\label{eq:cool_power}
\end{equation}
\begin{equation}
T_{i,t}^{\mathrm{in}} - T_{i,t-1}^{\mathrm{in}} =
\frac{P_{i,t}^{\mathrm{IT}} + P_{i,t}^{\mathrm{Aux}} - H_{i,t}}{\rho_o \varepsilon f_i}
+ \left(T_{i,t}^{\mathrm{out}} - T_{i,t-1}^{\mathrm{in}}\right)\left(1 - e^{-\kappa}\right)
\label{eq:indoor_temp_dynamics}
\end{equation}
\begin{equation}
P_{i}^{\mathrm{Cmin}} \leq P_{i,t}^{\mathrm{C}} \leq P_{i}^{\mathrm{Cmax}}
\label{eq:cool_power_bounds}
\end{equation}
\begin{equation}
T_{i}^{\mathrm{min}} \leq T_{i,t}^{\mathrm{in}} \leq T_{i}^{\mathrm{max}}
\label{eq:indoor_temp_bounds}
\end{equation}
where $H_{i,t}$ denotes the thermal cooling power of DC $i$ at time $t$, $\beta_i$ is the coefficient of performance (COP) of the cooling facility, and $P_{i}^{\mathrm{Cmin}}$ and $P_{i}^{\mathrm{Cmax}}$ are the minimum and maximum cooling power. $T_{i,t}^{\mathrm{in}}$ and $T_{i,t}^{\mathrm{out}}$ represent the indoor and outdoor temperatures of DC $i$ at time $t$, $T_{i}^{\mathrm{min}}$ and $T_{i}^{\mathrm{max}}$ denote the acceptable indoor temperature range for safe IT operation. The parameter $\rho_o$ is the air density, $\varepsilon$ is the specific heat capacity of air, $f_i$ denotes the airflow rate, and $\kappa$ is the thermal insulation coefficient.  
Eq.\eqref{eq:cool_power} describes the relationship between cooling power consumption and the effective thermal cooling output.  The indoor temperature dynamics are captured by Eq.\eqref{eq:indoor_temp_dynamics}, which balances heat generation from IT and auxiliary systems with cooling removal and thermal exchange with the ambient environment \cite{wahlroosFutureViewsWaste2018}. Constraints~\eqref{eq:cool_power_bounds}–\eqref{eq:indoor_temp_bounds} ensure the cooling power and indoor temperature limits.

\subsubsection{Workload Management Model}
\label{workload}
$A_{i,t}^{I}$ and $A_{i,t}^{B}$ are denoted as the forecasted arrival amounts of IWs and BWs of DC $i$ at time $t$. Correspondingly, $w_{i,t}^{I}$ and $w_{i,t}^{B}$ represent the processed amounts of IWs and BWs of DC $i$ at time $t$. The total workload processed by DC $i$ at time $t$ is expressed as \eqref{eq:total_workload}.

BWs are delay-tolerant and can be flexibly scheduled over time, but each BW must be completed before its individual deadline, as formulated in 
\eqref{eq:deadline_constraint}-\eqref{eq:batch_workload_feasibility}.
In contrast, IWs are latency-sensitive and should be processed within stringent response-time requirements, rather than being deferred across time periods. Therefore, in the baseline model, the locally processed IWs should satisfy \eqref{eq:interactive_workload}.
\begin{equation}
w_{i,t} = w_{i,t}^{\mathrm{I}} + w_{i,t}^{\mathrm{B}}.
\label{eq:total_workload}
\end{equation}
\begin{equation}
G_{i,t} = \sum_{\tau=1}^{t} A^\mathrm{B}_{i,\tau},
\label{eq:arrival_accumulation}
\end{equation}
\begin{equation}
D_{i,t} = \sum_{\tau=1}^{t} \delta_{i,\tau}(t) A_{i,\tau}^{\mathrm{B}}.
\label{eq:deadline_constraint}
\end{equation}
\begin{equation}
\delta_{i,\tau}(t) = 
\begin{cases}
1, & \text{if } T_{i,\tau}^\mathrm{d} \leq t, \\
0, & \text{if } T_{i,\tau}^\mathrm{d} > t.
\end{cases}
\label{eq:batch_deadline_indicator}
\end{equation}
\begin{equation}
D_{i,t} \leq \sum_{\tau=1}^{t} w_{i,\tau}^{\mathrm{B}} \leq G_{i,t},
\label{eq:batch_workload_feasibility}
\end{equation}
\begin{equation}
w_{i,t}^{\mathrm{I}} = A_{i,t}^\mathrm{I},
\label{eq:interactive_workload}
\end{equation}
where $G_{i,t}$ denotes the cumulative arrival of BWs up to time $t$, $D_{i,t}$ denotes the cumulative number of BWs that must be completed before time $t$, the binary indicator $\delta_{i,\tau}(t)$ determines whether a BW must be completed at or before time $t$, $T_{i,t}^d$ denotes the deadline for BWs arriving at time $t$.  The accumulated processing of BWs is limited by  $D_{i,t}$ and $G_{i,t}$.
\subsubsection{Active Server Number}
The total number of active servers is the sum of those handling IWs and BWs.
\begin{equation}
m_{i,t} = m_{i,t}^{\mathrm{IM}} + m_{i,t}^{\mathrm{BM}}
\label{eq:server_decomposition}
\end{equation}
where $m_{i,t}^{IM}$ and $m_{i,t}^{BM}$ denote the number of servers used to process IWs and BWs respectively.  

The relationship between the active servers and processed IWs is shown in \eqref{eq:server_allocation1}, where the M/M/1 queue model \cite{jiRobustOperationMinimizing2022} is applied to ensure QoS. and the relationship between the active servers and processed BWs is shown in \eqref{eq:server_allocation2}. The total number of active servers is constrained by the maximum available server capacity \eqref{eq:server_capacity}.
\begin{equation}
m_{i,t}^{\mathrm{IM}} = \frac{w_{i,t}^{\mathrm{I}}}{\gamma_i - 1/\nu_i}
\label{eq:server_allocation1}
\end{equation}
\begin{equation}
m_{i,t}^{\mathrm{BM}} = \frac{w_{i,t}^{\mathrm{B}}}{\gamma_i}
\label{eq:server_allocation2}
\end{equation}
\begin{equation}
m_{i,t} \leq {M}_i
\label{eq:server_capacity}
\end{equation}

\subsubsection{DC Operational Profit Model}
In this paper, we assume that each DC is fully powered by the power grid. Accordingly, the power balance can be expressed as \eqref{eq:dc_power_balance}.
\begin{equation}
P_{i,t}^{\text{DC}} 
= P_{i,t}^{\text{g}}, \quad \forall i \in \mathcal{I},\, 
\forall t \in \mathcal{T}.
\label{eq:dc_power_balance}
\end{equation}
\begin{equation}
0 \leq P^{\mathrm{g}}_{i,t} \leq {P}^{\mathrm{max}}_i, \quad \forall i \in \mathcal{I},\, 
\forall t \in \mathcal{T}.
\label{eq:dc_limit}
\end{equation}
where $P_{i,t}^{\text{g}}$ denotes the power purchased from the grid of DC $i$ at time $t$, should be subject to the physical limits \eqref{eq:dc_limit}. 

The total operation profit of each DC consists of three parts:  
(i) the revenue generated from receiving computation workloads from end-users $R_{i,t}$,  
(ii) the fixed equipment depreciation cost associated with task execution $C_{i,t}^{\mathrm{D}}$, (iii) the electricity purchase cost from the power grid $C_{i,t}^{\mathrm{G}}$.  

\begin{equation}
R_{i,t} = a_i\left(A^\mathrm{I}_{i,t} + A^\mathrm{B}_{i,t}\right)
\label{eq:R_def}
\end{equation}
\begin{equation}
C_{i,t}^{\mathrm{D}} = b_i\bigl(w^\mathrm{I}_{i,t} + w^\mathrm{B}_{i,t}\bigr)
\label{eq:Cdep_def}
\end{equation}
\begin{equation}
C_{i,t}^{\mathrm{G}} = \lambda^\mathrm{g}_{i,t}\,P^\mathrm{g}_{i,t}
\label{eq:Cgrid_def}
\end{equation}
where $a_i$ denotes the unit service fee charged to end-users, $b_i$ denotes the unit depreciation cost of computing equipment, and $\lambda^g_{i,t}$ denotes the LMP.

Given this context, each DC is aimed to seek to determine its optimal workload scheduling and electricity purchase scheme by solving the following individual optimization problem:
\begin{equation}
\begin{aligned}
\max  & \sum_{t \in \mathcal{T}} \left(R_{i,t} - C_{i,t}^{\mathrm{D}} - C_{i,t}^{\mathrm{G}}\right) \\
\text{s.t.} \quad  & (1)-(19) 
\end{aligned}
\label{eq:total_cost_dc}
\end{equation}
\subsection{Power System Pricing Model }
At the upper level, given the electricity demand of each DC, the power system operator determines the LMPs via solving the direct current optimal power flow. The objective function of the energy pricing model is to minimize the total generation cost, presented as:

\begin{equation}
\min \sum_{t \in \mathcal{T}} \sum_{g \in {\Omega}_G} \left( a_g (P^\mathrm{G}_{g,t})^2 + b_g P^\mathrm{G}_{g,t} \right)
\label{eq:objective}
\end{equation}
The objective is subject to the following constraints:
\begin{equation}
\sum_{g \in {\Omega}_G} P_{g,t}^{\mathrm{G}} +
\sum_{w \in {\Omega}_W} P_{w,t}^{\mathrm{W}} +
\sum_{s \in {\Omega}_{PV}} P_{s,t}^{\mathrm{PV}}
= \sum_{i \in \mathcal{I}} P_{i,t}^{g} :  \bar{\lambda}_t 
\label{eq:balance}
\end{equation}
\begin{equation}
\begin{aligned}
&- F_l \leq \;
\sum_{g \in \Omega_G} \pi_{gl} P_{g,t}^{\mathrm{G}}
+ \sum_{w \in \Omega_W} \pi_{wl} P_{w,t}^{\mathrm{W}}
+\\ \sum_{s \in \Omega_{PV}} &\pi_{sl} P_{s,t}^{\mathrm{PV}} 
- \sum_{i \in \mathcal{I}} \pi_{il} P_{i,t}^{g}
\leq F_l, \quad \forall l \in \mathcal{E}_L : \tau^+_{l,t}, \tau^-_{l,t}
\end{aligned}
\label{eq:lineflow}
\end{equation}
\begin{equation}
P_{g}^{\mathrm{Gmin}} \leq P_{g,t}^{\mathrm{G}} \leq P_{g}^{\mathrm{Gmax}}
\label{eq:thermal_limit}
\end{equation}
\begin{equation}
\begin{aligned}
0 \leq P_{w,t}^{\mathrm{W}} \leq P_{w,t}^{\mathrm{Wmax}}, \quad
0 \leq P_{s,t}^{\mathrm{PV}} \leq P_{s,t}^{\mathrm{PVmax}}
\label{eq:pv_limit}
\end{aligned}
\end{equation}
where ${\Omega}_G$, ${\Omega}_W$, ${\Omega}_{PV}$ are the sets of thermal, PV and wind generators, respectively, $a_g$ and $b_g$ denote the quadratic and linear cost coefficients of thermal generator $g$, $\pi_{gl}$, $\pi_{wl}$, $\pi_{sl}$, and $\pi_{il}$ denote the power transfer distribution factors (PTDFs) corresponding to each unit and load bus on line $l$ \cite{fangEvaluationLMPIntervals2016}; $F_l$ represents the thermal rating of line $l$.
Constraint~\eqref{eq:balance} guarantees the global power balance. Constraint~\eqref{eq:lineflow} represents the PTDF-based line flow constraints, ensuring that the power flow on each transmission line remains within its thermal rating $F_l$.   
Constraint~\eqref{eq:thermal_limit} and \eqref{eq:pv_limit} impose the generation capacity limits for the thermal units and renewable units. 

Given the Lagrangian function $\mathcal{L}$ of \eqref{eq:objective} that incorporates the global power balance \eqref{eq:balance} and PTDF-based line flow constraints \eqref{eq:lineflow}, the LMP of DC $i$ and time $t$ is defined as:
\begin{equation}
\lambda_{i,t} = \frac{\partial \mathcal{L}}{\partial P_{i,t}^{g}} 
= -\bar{\lambda}_t + \sum_{l \in \mathcal{E}_L} \pi_{il} \left( \tau^+_{l,t} - \tau^-_{l,t} \right)
\label{eq:LMP_derivation}
\end{equation}
where $\bar{\lambda}_t$ is the dual variable associated with the global power balance constraint ~\eqref{eq:balance} , and $\tau^+_{l,t}$ and $\tau^-_{l,t}$ are the dual variables corresponding to the upper and lower bounds of the line flow constraint for line $l$ \eqref{eq:lineflow}, respectively. The first term $-\bar{\lambda}_t$ represents the system-wide marginal electricity price, which is common to all nodes, while the second term accounts for the marginal congestion effect introduced by the transmission network through the PTDFs ($\pi_{il}$), which can explicitly capture the marginal impact of DC demand on line congestion and nodal prices.  
Therefore, $\lambda_{i,t}$ captures both the marginal cost of electricity generation and the marginal cost of network congestion, serving as the price signal fed back to the DC workload scheduling.

\section{Problem Reformulation with P2P-CSM}
\label{Problem Reformulation with P2P-CSM}
In this section, the proposed P2P-CSM and the reformulation of the optimization problem will be presented.
\subsection{Peer-to-Peer Cloud Service Market}
P2P-CSM provides a novel paradigm for decentralized workload management among geo-distributed DCs, enabling them to exchange cloud services to balance computational surpluses and deficits across regions. 
In this mechanism, cloud service trading essentially represents IWs' spatial migration; each DC acts as both a potential service provider and a service consumer, and can flexibly buy or sell computing resources to other DCs under bilateral contracts. Each transaction is characterized by two key parameters: the volume of cloud service and the corresponding service price. 

Let $w_{ij,t}^{}$ denote the volume of cloud service purchased by DC $i$ from DC $j\in \mathcal{I}_i $ at time $t$. A positive value ($w_{ij,t}^{} > 0$) indicates that DC $i$ buys cloud services from DC $j$, while a negative value ($w_{ij,t}^{} < 0$) implies that DC $i$ provides cloud services to DC $j$. Let $w_{ij}^{\max}$ denote the maximum amount of workload that can be migrated between DC $i$ and DC $j$ within a single time interval. The workload transactions must satisfy \eqref{eq:workload_balance} and \eqref{eq:tradinglimit}.
\begin{equation}
w_{ij,t}^{\mathrm{}} + w_{ji,t}^{\mathrm{}} = 0, 
\quad \forall i \in \mathcal{I},\, 
\forall j \in \mathcal{I}_i,\, 
\forall t \in \mathcal{T}
\label{eq:workload_balance}
\end{equation}
\begin{equation}
        -w_{ij}^{\max} \le w_{ij,t} \le w_{ij}^{\max}, \quad
    \forall i \in \mathcal{I},\, j \in \mathcal{I}_i,\, t \in \mathcal{T}
    \label{eq:tradinglimit}
\end{equation}

Let $\mu_{ij,t}$ denote the bilateral transaction price set by DC $i$ for DC $j$ at time $t$. For any successful trading pair, both parties must agree on a bilateral price consistency scheme, as enforced by \eqref{eq:price_agreement}. This scheme does not impose a uniform price across all trading pairs. Instead, it enforces price consistency only within each bilateral transaction, allowing different trading links to have different prices according to their local marginal values.
\begin{equation}
\mu_{ij,t} = \mu_{ji,t},
\quad \forall i \in \mathcal{I},\, 
\forall j \in \mathcal{I}_i,\, 
\forall t \in \mathcal{T}
\label{eq:price_agreement}
\end{equation}

After participating in the P2P-CSM, the total IWs of each DC is updated to include both the locally processed and transaction portions \eqref{eq:interactive_workload_new}. Furthermore, the total IWs in the entire system must satisfy the global balance \eqref{eq:interactive_balance_new}, which ensures that no workload is lost during P2P cloud service exchanges.
\begin{equation}
w_{i,t}^{\mathrm{I}} = A_{i,t}^{\mathrm{I}} - \sum_{j \in \mathcal{I}_i} w_{ij,t}^{\mathrm{I}},
\quad \forall i \in \mathcal{I},\, t \in \mathcal{T}
\label{eq:interactive_workload_new}
\end{equation}
\begin{equation}
\sum_{i \in \mathcal{I}} w_{i,t}^{\mathrm{I}}
= \sum_{i \in \mathcal{I}} A_{i,t}^{\mathrm{I}},
\quad \forall t \in \mathcal{T}
\label{eq:interactive_balance_new}
\end{equation}

Let $\bm{\mu}_{i,t}=\{ {\mu}_{ij,t} \}_{j \in \mathcal{I}_i} $ and
$\bm{w}_{i,t} = \{ {w}_{ij,t}^{} \}_{j \in \mathcal{I}_i}$
represent the trading price and volume vector between DC $i$ and its neighboring DCs at time $t$. By incorporating the P2P-CSM costs into the original profit function \eqref{eq:total_cost_dc}, the optimization problem of DC $i$ can be improved as \eqref{eq:total_cost_dc_new}.
\begin{equation}
\begin{aligned}
\max\;
& \sum_{t \in \mathcal{T}} 
\left(
R_{i,t} - C_{i,t}^{\mathrm{D}} - C_{i,t}^{\mathrm{G}}
- \bm{\mu}_{i,t}^{\top} \bm{w}_{i,t}
\right) \\
\text{s.t.} \quad 
 & (1) - (8) ,\; (10) - (19),\; (30) - (33)
\end{aligned}
\label{eq:total_cost_dc_new}
\end{equation}
Therefore, for the DC cluster with P2P-CSM, the multi-DCs optimization problem can be equivalently represented by the following social welfare maximization problem \eqref{eq:total_cost_dc_all}.
\begin{equation} 
\begin{aligned}
\max
& \sum_{i \in \mathcal{I}} \sum_{t \in \mathcal{T}}
\left(
R_{i,t} - C_{i,t}^{\mathrm{D}} - C_{i,t}^{\mathrm{G}}
- \bm{\mu}_{i,t}^{\top} \bm{w}_{i,t}
\right) \\
\text{s.t.} \quad 
 & (1) - (8) ,\; (10) - (19),\; (30) - (34)
\end{aligned}
\label{eq:total_cost_dc_all}
\end{equation}


\subsection{Problem reformulation with fair pricing scheme}
Essentially, the social welfare problem \eqref{eq:total_cost_dc_all} is the centralized equivalent of the decentralized equilibrium in the P2P-CSM. However, the absence of a central coordinator and the requirement for autonomous decision-making render a direct centralized solution impractical. Consequently, the global problem is decomposed into $I$ local subproblems, where each DC independently optimizes its workload scheduling and power procurement strategy, as formulated in \eqref{eq:total_cost_dc_new}.

Nevertheless, these individual subproblems are interdependent. The decision variables, specifically the trading prices $\bm{\mu}_{i,t}$ and volume $\bm{w}_{i,t}$, are inherently coupled across neighboring DCs through workload balance constraint \eqref{eq:workload_balance} and price agreement constraint \eqref{eq:price_agreement}. 
To capture these coupling relationships more compactly, let $\mathcal{E}$ denote the set of all trading pairs $(i,j)$ among $I$ DCs, such that $|\mathcal{E}| = I(I-1)/2$. Accordingly, the coupling constraint \eqref{eq:workload_balance} can be rewritten in a compact form as:
\begin{equation}
\sum_{i\in\mathcal{I}} \bm{E}_i \bm{w}_{i,t} = \bm{0},
\quad \forall\, t \in \mathcal{T}
\label{eq:compact_coupling}
\end{equation}
where $\bm{E}_i \in \mathbb{R}^{|\mathcal{E}| \times (I-1)}$ is a mapping matrix from nodes to links.  
For a fixed and full interconnection P2P topology, $\bm{E}_i$ is predefined as a time-invariant and sparse node-to-link incidence matrix with exactly $I-1$ non-zero entries. The number of its non-zero entries equals the degree of node $i$, denoted by $\mathcal{I}_i$. Specifically, if a non-zero entry exists at the $m$-th row and $k$-th column of $\bm{E}_i$, it signifies that the $k$-th local trading partner of DC $i$ is associated with the $m$-th global edge $(i, j) \in \mathcal{E}$.

Next, let $\bm{\mu}_{t} = \{ \mu_{(i,j),t} \}_{(i,j) \in \mathcal{E}} \in \mathbb{R}^{|\mathcal{E}|}$ denote the vector of Lagrange multipliers associated with \eqref{eq:compact_coupling}. Each component $\mu_{(i,j),t}$ serves as the shadow price for the bilateral cloud service exchange across link $(i,j)$ at time $t$. Since the bilateral price agreement in \eqref{eq:price_agreement} requires the two directional prices of each trading pair to be identical, we introduce $(\mu_{(i,j),t})$ as the common edge-based price. Accordingly, the directional price variables can be rewritten as
\begin{equation}
\mu_{ij,t} = \mu_{ji,t} := \mu_{(i,j),t}, \quad \forall, (i,j) \in \mathcal{E},, \forall, t \in \mathcal{T}
\end{equation}
By exploiting this symmetry, the local price vector $\bm{\mu}_{i,t}$ can be expressed as a linear mapping of the global price vector: 
\begin{equation}
\bm{E}_i^{\top} \bm{\mu}_{t} = \bm{\mu}_{i,t}
\label{eq:price_mapping}
\end{equation}
Substituting this into \eqref{eq:total_cost_dc_new}, the reformulated local optimization problem for DC $i$ is obtained as:

\begin{equation}
\begin{aligned}
\max\;
& \sum_{t \in \mathcal{T}} 
\Bigl(
R_{i,t} - C_{i,t}^{\mathrm{D}} - C_{i,t}^{\mathrm{G}}
- \bm{\mu}_{t}^{\top} \bm{E}_i \bm{w}_{i,t}
\Bigr) \\
\text{s.t.}\quad 
 (1) - &(8),\; (10) - (19),\; (30) - (34), \; (37-39).
\end{aligned}
\label{eq:total_cost_dc_new_refined}
\end{equation}

In this reformulation, the original coupling constraints \eqref{eq:workload_balance} and \eqref{eq:price_agreement} are effectively decoupled through the proposed fair pricing mechanism and Lagrangian duality. Specifically, the workload balance condition \eqref{eq:workload_balance} is internalized into the individual objectives as a penalty term involving the global prices, while the price agreement condition \eqref{eq:price_agreement} is inherently satisfied by the consensus on the global price vector $\bm{\mu}_{t}$. Consequently, given a fixed $\bm{\mu}_{t}$, each DC can independently optimize its local objective \eqref{eq:total_cost_dc_new_refined}.
However, $\bm{\mu}_{t}$ serves as a global dual vector that necessitates  a central coordinator for updates, which contradicts the fully decentralized, and retaining dual variables in the problem brings challenges in dealing with the bilinear term $\bm{\mu}_{t}^{\top} \bm{E}_i \bm{w}_{i,t}$. These challenges will be systematically addressed in Section~\ref{solution}.

\section{Solution Methodology}
\label{solution}
In this section, a dual consensus ADMM-based decentralized algorithm is proposed to solve the P2P cloud service trading problem, and a bisection-assisted iterative algorithm is introduced for the bi-level computation-electricity coordination interaction.
 
\subsection{Dual Consensus ADMM-based Decentralized Algorithm}
The presence of the global variable $\bm{\mu}_{t}$ and the bilinear coupling term $\bm{\mu}_{t}^{\top} \bm{E}_i \bm{w}_{i,t}$ renders conventional ADMM algorithms inapplicable. Inspired by \cite{changMultiAgentDistributedOptimization2015}, we propose a dual consensus ADMM-based decentralized algorithm to overcome these obstacles,


First, for simplicity,  let $\bm{x}_{i,t}$ denote the set of all local decision variables for DC $i$ at time $t$, which encompasses the cloud service trading vector $\bm{w}_{i,t}$ and other internal operational variables.
The lagrange multipliers ${{\bm{\mu}}_{i,t}}$ associated with \eqref{eq:compact_coupling} represent the shadow prices of the P2P transactions. By the Lagrangian dual decomposition, we could define the local dual function at \eqref{eq:dual_form}.
\begin{equation}
\min_{\bm{\mu}_{t}}
\;\sum_{t\in\mathcal{T}}
\varphi_i(\bm{\mu}_{t})
\label{eq:dual_form}
\end{equation}
\begin{equation}
\varphi_i({{\bm{\mu}}_{t}})
=
\max_{\bm{x}_{i,t}}
\left\{
 R_{i,t}-C_{i,t}^{\mathrm{D}}-C_{i,t}^{\mathrm{G}}
-{\bm{\mu}}_{t}^{\top}E_i\bm{w}_{i,t}
\right\}
\label{eq: dual_term}
\end{equation}

In a decentralized environment, the global price vector $\bm{\mu}_t$ is not directly accessible. Instead, each DC $i \in \mathcal{I}$ maintains a local pricing estimate, denoted by $\hat{\bm{\mu}}_{i,t}$, to represent its perception of the global market state. Within the P2P communication network, each DC $i$ receive estimates from neighbors ${\hat{\bm{\mu}}_{j,t}},j \in \mathcal{I}_i$. To obtain global optimality, the price agreement, namely the dual consensus, must be achieved. With the assistance of slack variables $\bm{s}_{ij,t}$, the dual consensus is expressed as follows:
\begin{equation} 
\hat{\bm{\mu}}_{i,t} = \bm{s}_{ij,t},\quad
\hat{\bm{\mu}}_{j,t} = \bm{s}_{ij,t},\quad
\forall (i,j) \in \mathcal{E}, \forall t \in \mathcal{T} 
\label{eq:consensus_constraints} 
\end{equation}

For the dual problem \eqref{eq: dual_term} with constrains \eqref{eq:consensus_constraints}, According to \cite{mateosDistributedSparseLinear2010}, ADMM leads to the following iterative updates to get $\hat{\bm{\mu}}_{i,t}$:

\begin{equation}
\begin{aligned}
\bm{z}_{i,t}^{k}
= \bm{z}_{i,t}^{\,k-1}
  + \rho \sum_{j \in \mathcal{I}_i}
      \big(\hat{\bm{\mu}}_{i,t}^{k}-\hat{\bm{\mu}}_{j,t}^{k}\big)
\end{aligned}
\label{eq:ax}
\end{equation}
\begin{equation}
\begin{aligned}
\hat{\bm{\mu}}_{i,t}^{k}&=\arg \ \min_{\hat{\bm{\mu}}_{i,t}}  \;
\sum_{t \in \mathcal{T}} \Bigg(
\varphi_i\!\left(\hat{\bm{\mu}}_{i,t}\right)
   + \hat{\bm{\mu}}_{t}^{\top}\bm{z}_{i,t}
 \\&  + \rho \sum_{j \in \mathcal{I}_i}
     \bigg\| \hat{\bm{\mu}}_{i,t}
     - \tfrac{1}{2}\big(\hat{\bm{\mu}}_{i,t}^{k-1}+\hat{\bm{\mu}}_{j,t}^{k-1}\big)
     \bigg\|_2^{2}
\Bigg)
\end{aligned}
\label{eq:minmax}
\end{equation}
where $\bm{z}_{i,t}$ is the auxiliary variable, $\rho$ is the given penalty parameter. 

For \eqref{eq:minmax}, by substituting $\varphi_i(\hat{\bm{\mu}}_{i,t})$ in \eqref{eq: dual_term} into \eqref{eq:minmax}, we can observe that the local update at each DC $i$ is, in fact, a min-imax optimization problem:

\begin{equation}
\begin{aligned}
\hat{\bm{\mu}}_{i,t}^{k}=&\arg  \min_{\hat{\bm{\mu}}_{i,t}} \max_{\bm{x}_{i,t}} \;
\sum_{t \in \mathcal{T}} \Bigg(
    R_{i,t}
   - C_{i,t}^{\mathrm{D}}
   - C_{i,t}^{\mathrm{G}}
   - \hat{\bm{\mu}}_{i,t}^{\top}\bm{E}_{i}\bm{w}_{i,t}
\\
   &   + \hat{\bm{\mu}}_{i,t}^{\top}\bm{z}_{i,t} + \rho \sum_{j \in \mathcal{I}_i}
     \bigg\| \hat{\bm{\mu}}_{i,t}
     - \tfrac{1}{2}\big(\hat{\bm{\mu}}_{i,t}^{k-1}+\hat{\bm{\mu}}_{j,t}^{k-1}\big)
     \bigg\|_2^{2}
\Bigg)
\end{aligned}
\label{eq:minmax2}
\end{equation}
It is observed that the function \eqref{eq:minmax2} is convex in $\hat{\bm{\mu}}_{i,t}$ with given $\bm{x}_{i,t}$,  and concave in $\bm{x}_{i,t}$ with given $\hat{\bm{\mu}}_{i,t}$. Thus, according to the min-max theory~\cite{bertsekasConvexAnalysisOptimization2003}, the optimal solution is exactly the saddle point. By completing the quadratic term, the primal and dual variables are solved as follows:

\begin{equation}
\begin{aligned}
\bm{x}_{i,t}^{k}
= \arg\min_{\bm{x}_{i,t}^{k}} &\Bigg(
 - R_{i,t}
   + C_{i,t}^{\mathrm{D}}
   + C_{i,t}^{\mathrm{G}}
   + \frac{\rho}{4\lvert \mathcal{I}_i\rvert}
     \bigg\|
        \rho^{-1}\bm{E}_i\,\bm{w}_{i,t}
       \\
      & - \rho^{-1}\bm{z}_{i,t}^{\,k-1}
        + \!\!\sum_{j \in \mathcal{I}_i}\!\!
          \big(\hat{\bm{\mu}}_{i,t}^{\,k-1}+\hat{\bm{\mu}}_{j,t}^{\,k-1}\big)
     \bigg\|_2^{2}
\Bigg)
\end{aligned}
\label{eq:pm}
\end{equation}
\begin{equation}
\begin{aligned}
\hat{\bm{\mu}}_{i,t}^{k}
= \frac{1}{2\lvert \mathcal{I}_i\rvert}
\Bigg(
     \sum_{j \in \mathcal{I}_i}
       \big(\hat{\bm{\mu}}_{i,t}^{\,k-1}+\hat{\bm{\mu}}_{j,t}^{\,k-1}\big)
     - \rho^{-1}\big(\bm{z}_{i,t}^{\,k-1} - \bm{E}_i\,\bm{w}_{i,t}^{k}\big)
\Bigg)
\end{aligned}
\label{eq:du}
\end{equation}

To be clearer, Algorithm \ref{alg:decentralized} summarizes the proposed decentralized algorithm. The cloud service trading and mutual payment are attained simultaneously by this holistic algorithm, and the only exchanged information are local estimates of global prices.
\begin{algorithm}[t]
\caption{Dual Consensus ADMM-Based Decentralized Algorithm.}
\label{alg:decentralized}
\begin{algorithmic}[1]
\STATE \textbf{Initialization}: set iteration time $k = 1$, $\hat{\bm{\mu}}_{i,t}= \bm{0}$, $\bm{z}_{i,t}= \bm{0}$ for DC $i \in \mathcal{I}$, given parameter $\rho$;
\REPEAT
    \FOR{each DC $i \in \mathcal{I}$ \textbf{(in parallel)}}
        \STATE Update primal variables $\bm{x}_{i,t}^{k}$  according to \eqref{eq:pm};
        \STATE Update dual variables $\hat{\bm{\mu}}_{i,t}^{k}$ according to \eqref{eq:du};
        \STATE Exchange $\hat{\bm{\mu}}_{i,t}^{k}$ with neighboring DCs $j \in \mathcal{I}_{i}$,
        and receive $\hat{\bm{\mu}}_{j,t}^{k}$ from neighbors;
        \STATE Update auxiliary variables $\bm{z}_{i,t}^{k}$ according to \eqref{eq:ax};
    \ENDFOR
    \STATE $k = k + 1$;
\UNTIL{a predefined convergence criterion is satisfied}
\end{algorithmic}
\end{algorithm}
\subsection{Bisection-Assisted Iterative Algorithm}
In the proposed bi-level computation-electricity coordination framework, the upper- and lower-level problems are iteratively solved to achieve convergence. Although the Karush-Kuhn-Tucker reformulation may provide an exact centralized solution, it would require private information from all DCs and introduce complex complementarity constraints, which is inconsistent with the decentralized P2P market design. At the upper level, the power grid determines the LMPs based on generation dispatch and network constraints. At the lower level, each DC responds to these price signals by optimizing its workload allocation and energy consumption. The resulting power demand profiles are subsequently fed back to the upper level to update LMPs for the following iteration. This bidirectional iterative interaction persists until convergence is reached. 
The convergence criterion is defined as
\begin{equation}
\left\| P_{i,t}^{\mathrm{g},k} - P_{i,t}^{\mathrm{g},k-1} \right\|_2 \leq \xi,
\quad \forall i \in \mathcal{I},\, \forall t \in \mathcal{T}
\label{eq:power_change_limit}
\end{equation}
where $\xi$ denotes the convergence tolerance and $k$ is the iteration time.

However, due to the strong coupling between two levels, oscillations may occur during the iterative process. 
Specifically, the power consumption of DCs may fluctuate between two previous states, expressed as
\begin{equation}
P_{i,t}^{\mathrm{g},k} = P_{i,t}^{\mathrm{g},k-2}, 
\quad \forall i \in \mathcal{I},\, \forall t \in \mathcal{T}
\label{eq:oscillation}
\end{equation}

To address oscillations and ensure stable convergence during iteration, a bisection-assisted iterative algorithm is proposed. The algorithm embeds a bisection-based mechanism into the bi-level framework to dynamically construct and update an operation interval for each DC once oscillation is detected. This interval adaptively bounds the feasible operating region of each DC and is gradually narrowed to guide this process toward convergence.

Specifically, if oscillation occurs at the $k$-th iteration, the power demand of DC~$i$ at time~$t$ is denoted as $P_{i,t}^{\mathrm{g},k}$. 
An operational interval $\bigl[P_{i,t}^{\min}, P_{i,t}^{\max}\bigr]$, which represents the feasible operation range of DC~$i$ could be constructed based on the results of the two most recent iterations:
\begin{equation}
\begin{aligned}
&P_{i,t}^{\min} = \min(P_{i,t}^{\mathrm{g},k}, P_{i,t}^{\mathrm{g},k-1}) \\
&P_{i,t}^{\max} = \max(P_{i,t}^{\mathrm{g},k}, P_{i,t}^{\mathrm{g},k-1})
\end{aligned}
\label{eq:interval}
\end{equation}

\begin{algorithm}[t]
\caption{Bisection-Assisted Iterative Algorithm}
\label{alg:dual_boundary}
\begin{algorithmic}[1]
\STATE \textbf{Initialization}: set iteration time $k = 1$, convergence tolerances $\xi > 0$;
\REPEAT
    \STATE Solve the upper level power system pricing problem to update the LMPs;
    \STATE Solve the lower level DC energy and workload management problem to update the power demand $P_{i,t}^{\mathrm{g}}$;
    \IF{$P_{i,t}^{\mathrm{g},k}  = P_{i,t}^{\mathrm{g},k-2} $}
        \STATE \hspace{0.4cm}${P}_{i,t}^{\mathrm{g,min}} = \min(P_{i,t}^{\mathrm{g},k}, P_{i,t}^{\mathrm{g},k-1}),$
        \STATE \hspace{0.4cm}${P}_{i,t}^{\mathrm{g,max}} = \max(P_{i,t}^{\mathrm{g},k}, P_{i,t}^{\mathrm{g},k-1});$
        \STATE\hspace{0.4cm}$\hat{P}_{i,t}^{\mathrm{g}} = \left({P}_{i,t}^{\mathrm{g,min}} + {P}_{i,t}^{\mathrm{g,max}}\right)/2;$
        \STATE Solve the upper level problem based on $\hat{P}_{i,t}^{\mathrm{g}}$;
        \STATE Solve the lower level problem to obtain an auxiliary DC power consumption $\widetilde{P}_{i,t}^{\mathrm{g}}$;
        \STATE \hspace{0.4cm}$ {P}_{i,t}^{\mathrm{g,min}} = (\widetilde{P}_{i,t}^{\mathrm{g}} > \hat{P}_{i,t}^{\mathrm{g}})
        \hat{P}_{i,t}^{\mathrm{g}} + (\widetilde{P}_{i,t}^{\mathrm{g}} \le \hat{P}_{i,t}^{\mathrm{g}})
        {P}_{i,t}^{\mathrm{g,min}},$
        \STATE \hspace{0.4cm}$ {P}_{i,t}^{\mathrm{g,max}} = (\widetilde{P}_{i,t}^{\mathrm{g}} < \hat{P}_{i,t}^{\mathrm{g}})
       \hat{P}_{i,t}^{\mathrm{g}} + (\widetilde{P}_{i,t}^{\mathrm{g}} \ge \hat{P}_{i,t}^{\mathrm{g}})
       {P}_{i,t}^{\mathrm{g,max}};$
        \STATE Add constraint for DC $i$: ${P}_{i,t}^{\mathrm{g,min}} \le P_{i,t}^{\mathrm{g}} \le {P}_{i,t}^{\mathrm{g,max}}.$
    \ENDIF
    \STATE Set $k = k + 1;$
\UNTIL{convergence criterion \eqref{eq:power_change_limit} is satisfied.}
\end{algorithmic}
\end{algorithm}
The midpoint of this interval is selected as the tentative operating point for the operation range update:
\begin{equation}
\hat{P}_{i,t}^{\mathrm{g}} = \left({P_{i,t}^{\mathrm{g},\min} + P_{i,t}^{\mathrm{g},\max}}\right)/2
\label{eq:midpoint}
\end{equation}
Given the adjusted demand $\hat{P}_{i,t}^{\mathrm{g},}$, the upper-level problem is re-solved to update the auxiliary LMPs, and the corresponding lower-level optimization of each DC is performed to obtain an auxiliary power consumption $\widetilde{P}_{i,t}^{\mathrm{g}}$. 
The operational bounds are then updated according to the relative position between $\widetilde{P}_{i,t}^{\mathrm{g}}$ and $\hat{P}_{i,t}^{\mathrm{g}}$:
\begin{equation}
\begin{aligned}
 {P}_{i,t}^{\mathrm{g,min}} &= (\widetilde{P}_{i,t}^{\mathrm{g}} > \hat{P}_{i,t}^{\mathrm{g}})\,\hat{P}_{i,t}^{\mathrm{g}} 
 + (\widetilde{P}_{i,t}^{\mathrm{g}} \le \hat{P}_{i,t}^{\mathrm{g}})\,{P}_{i,t}^{\mathrm{g,min}} \\
 {P}_{i,t}^{\mathrm{g,max}} &= (\widetilde{P}_{i,t}^{\mathrm{g}} < \hat{P}_{i,t}^{\mathrm{g}})\,\hat{P}_{i,t}^{\mathrm{g}} 
 + (\widetilde{P}_{i,t}^{\mathrm{g}} \ge \hat{P}_{i,t}^{\mathrm{g}})\,{P}_{i,t}^{\mathrm{g,max}}
\end{aligned}
\label{eq:bound_update}
\end{equation}

Accordingly, the operational constraint for each DC is narrowed to:
\begin{equation}
{P}_{i,t}^{\mathrm{g,min}} \le P_{i,t}^{\mathrm{g}} \le {P}_{i,t}^{\mathrm{g,max}}
\label{eq:constraint_update}
\end{equation}
By iteratively updating the interval bounds based on the new auxiliary solutions, the feasible region for each DC is progressively contracted until stabilization is achieved. This algorithm effectively solves oscillations between iterations and ensures robust convergence of the computation-electricity coordination framework.

\section{Case Studies}
\label{Case Studies}
To evaluate the effectiveness of the proposed P2P-CSM, numerical simulation results are presented in this section.

\subsection{System Setup}
\begin{figure}[t]
    \centering
    \includegraphics[scale=0.8]{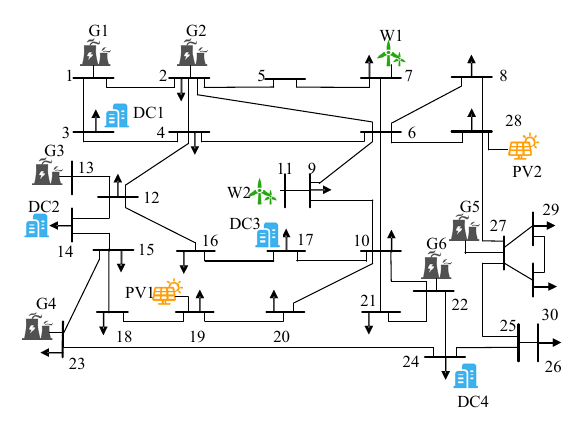}
        \vspace{-0.4cm}
    \caption{Modified IEEE 30-bus system with four DCs.}
    \vspace{-0.4cm}
    \label{fig:IEEE 30}
\end{figure}

\begin{figure}[t]
\centering
\includegraphics[scale =0.9]{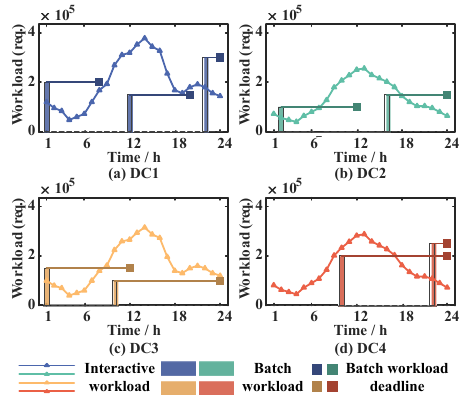} 
\vspace{-0.2cm}
\caption{Workload profiles of four DCs.}
\vspace{-0.5cm}
\label{workload_figure_ori}
\end{figure}

We conduct the case study on a modified IEEE 30-bus system comprising four thermal generators, two wind farms, and two PV farms. Four DCs are located at buses 4, 13, 17, and 24, respectively, as illustrated in Fig.~\ref{fig:IEEE 30}. All simulations are performed over a 24-hour scheduling horizon with a one-hour resolution. All computational experiments are implemented in MATLAB using the YALMIP modeling toolbox and the Gurobi solver.
The IWs and BWs profiles of DCs are shown in Fig. \ref{workload_figure_ori} \cite{zhouDataCenterLoad2025}. 
The operational parameters of the DCs follow \cite{chenInternetDataCenter2021}.
To reflect heterogeneity in cooling system, the COPs of the four DCs are assumed to differ as $[ \beta_1, \beta_2,\beta_3, \beta_4]=[6, 3, 3, 6]$. The indoor operating temperature for all DCs is set at 25°C. The parameters of thermal generators are from \cite{chaibOptimalPowerFlow2016}. 
Wind and PV generation profiles, as well as outdoor temperature data, are sourced from the National Renewable Energy Laboratory (NREL) \cite{NRELMeasurementInstrumentation} and the Hong Kong Observatory (HKO) \cite{RegionalWeatherHong}. 
The penalty parameter of the dual consensus-ADMM algorithm is set to $\rho = 4.5$ and the stopping criterion is dual residual $\left\| \hat{\bm{\mu}}_{i,t}^{k} - \hat{\bm{\mu}}_{i,t}^{k-1} \right\|_2
\le 0.02 \times 10^{-3}$. 

\vspace{-0.5cm}

\subsection{Convergence Analysis of the Computation-Electricity Coordination Framework}

\begin{figure}[h]
\centering
\includegraphics[scale=0.75]{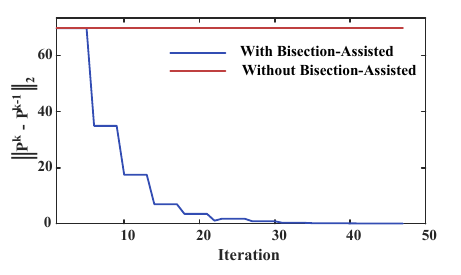}
\vspace{-0.2cm}
\caption{Iteration process of power residual \eqref{eq:power_change_limit}.}
\vspace{-0.5cm}
\label{Iteration process of power residual}
\end{figure}

We first examine the convergence behavior of the proposed bi-level computation-electricity coordination framework.
Fig. \ref{Iteration process of power residual} compares the power residual trajectories with and without the proposed bisection-assisted iterative algorithm.
Without bisection assistance, the residual remains at a persistently high level throughout the iterations, indicating that the interaction between the upper and lower levels oscillates.
In contrast, when the bisection-assisted step-size adjustment is activated, the residual decreases rapidly and falls below the prescribed tolerance within 47 iterations.

\begin{figure}[t]
\centering
\includegraphics[scale=0.75]{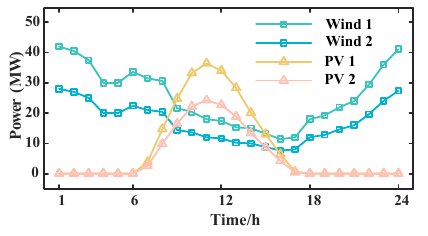}
\vspace{-0.2cm}
\caption{Power generation of wind and PV farms.}
\vspace{-0.2cm}
\label{Power generation of wind and PV farms}
\end{figure}

\begin{figure}[t]
\centering
\includegraphics[scale =0.75]{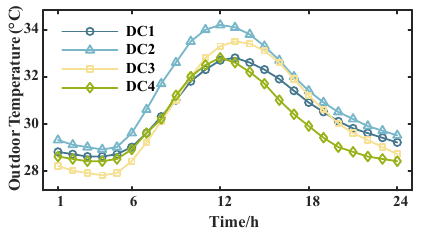}
\vspace{-0.2cm}
\caption{Outdoor temperature of four DCs.}
\label{Outdoor temperature}
\vspace{-0.5cm}
\end{figure}

\begin{figure}[b]
\centering
\vspace{-0.4cm}
\includegraphics[scale=0.75]{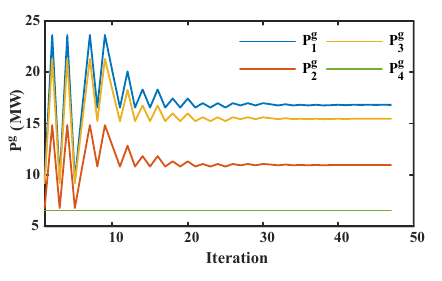}
\vspace{-0.2cm}
\caption{Iteration process of $P^\mathrm{g}_i$ at $t = 2$.}
\label{oscillation}
\end{figure}

Fig.\ref{oscillation} shows the iterative trajectories of the grid power $P_{i,t}^{\mathrm{g},k}$ for the four DCs at a representative time slot $t=2$. 
At the third iteration, the grid power returns to its value in the first iteration, indicating the occurrence of oscillation.
During the early iterations,  $P_{i,t}^{\mathrm{g},k}$ exhibits strong fluctuations because the DCs repeatedly update their workload scheduling and electricity demand in response to newly posted LMPs, while the power grid simultaneously updates the LMPs based on the changing injections.
With the proposed bisection-assisted iterative algorithm, oscillations are gradually narrowed. In total, 11 oscillation cycles are observed. All power variables ultimately converge to stable values after 47 iterations.
This clear performance demonstrates that the bisection-assisted iterative algorithm is essential for addressing oscillatory behavior and enabling reliable convergence to the electricity–computation coordination framework.

\vspace{-0.3cm}

\begin{figure}[t]
\centering
\includegraphics[scale=0.8]{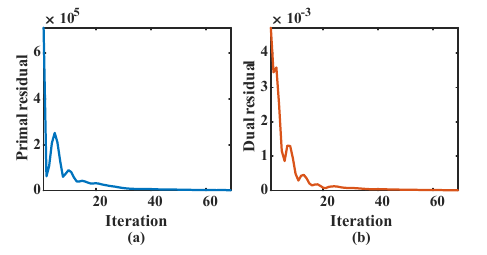}
\vspace{-0.2cm}
\caption{Dual consensus-ADMM iteration results.}
\vspace{-0.4cm}
\label{Dual consensus-ADMM iteration results}
\end{figure}

\begin{figure}[b]
\centering
\vspace{-0.6cm}
\includegraphics[scale=0.8]{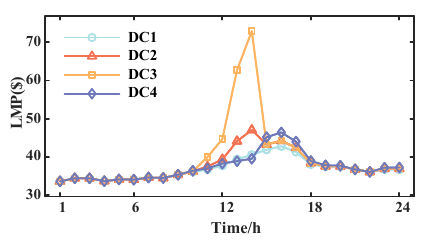}
\vspace{-0.3cm}
\caption{LMP without P2P-CSM.}
\vspace{-0.3cm}
\label{LMP without P2P-CSM}
\end{figure}

\begin{figure}[b]
\centering
\vspace{-0.2cm}
\includegraphics[scale=0.8]{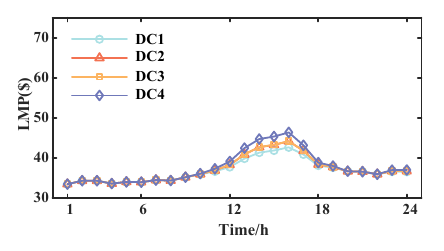}
\vspace{-0.3cm}
\caption{LMP with P2P-CSM.}
\label{LMP with P2P-CSM}
\end{figure}
\vspace{-0.8cm}
\subsection{Performance Evaluation of the P2P-CSM}
In each iteration of the bi-level coordination, the lower-level P2P-CSM problem is solved using the dual consensus ADMM-based decentralized algorithm. Fig. \ref{Dual consensus-ADMM iteration results} shows the primal and dual residuals in the final iteration, both of which converge within 68 iterations. We next analyze the trading behavior in the P2P-CSM, which is primarily driven by two factors: LMP variations and cooling efficiency differences.

\subsubsection{LMP-driven cloud service trading}

\begin{figure}[t]
\centering
\includegraphics[scale=0.8]{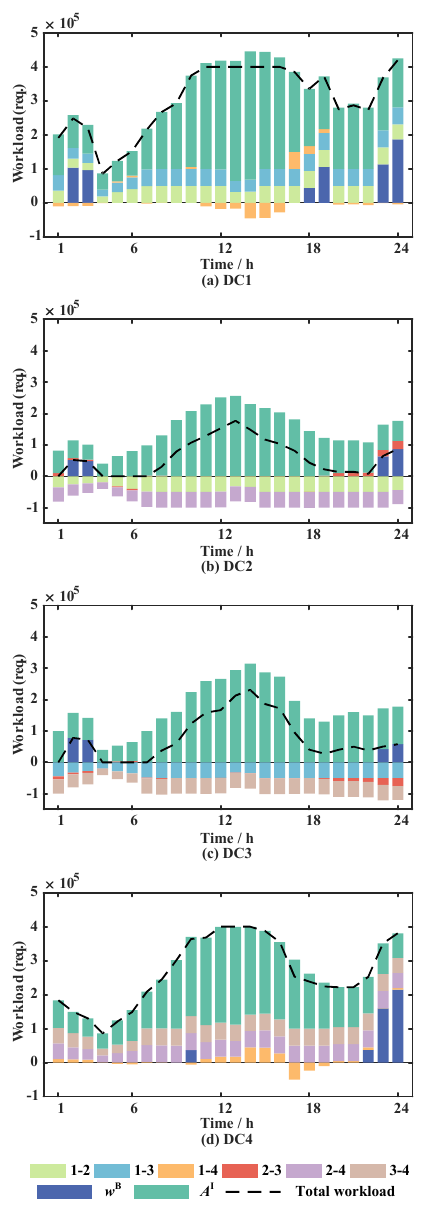}
\vspace{-0.2cm}
\caption{Workload profiles of four DCs with P2P-CSM.}
\label{Workload profiles of 4 DCs with P2P-CSM}
\vspace{-0.5cm}
\end{figure}

LMPs constitute a primary driver of cloud service trading in the proposed P2P-CSM. Fig.~\ref{LMP without P2P-CSM} presents the LMP trajectories without the P2P-CSM, while Fig.~\ref{LMP with P2P-CSM} shows the corresponding results when the market is enabled. During hours 11–18, substantial LMP disparities emerge across the four DCs. In the absence of the P2P-CSM, DC2 and DC3 experience pronounced price spikes, whereas DC1 maintains a much lower price level. These LMP differences create strong incentives for high price DCs to purchase cloud services from lower price peers to reduce their operating costs. This price-driven behavior is confirmed by the workload profiles in Fig.~\ref{Workload profiles of 4 DCs with P2P-CSM}. During hours 11–14, both DC2 and DC3 offload part of their IWs to DC1 and DC4, fully aligned with the economic incentives implied by the LMP differentials.

A noteworthy phenomenon appears during hours 15–16. Although DC1 still faces a lower LMP than DC4, Fig.~\ref{workload_figure_ori} shows that DC1 purchases cloud services from DC4. This behavior is attributable to physical limits on computational capacity. From hours 11–16, the system experiences its highest overall workload, and DC1 operates at its maximum processing capacity. As a result, DC1 can no longer accommodate additional workload transfers and must temporarily offload part of its own workload to DC4, despite facing a slightly lower LMP.

\subsubsection{Cooling efficiency-driven cloud service trading}
Cooling efficiency constitutes another key driver of cloud service trading in the proposed P2P-CSM. Even when electricity prices are identical across DCs, differences in cooling efficiency imply that processing the same amount of IWs results in different electricity consumption levels.

During hours 1–9 and 20–24, the system load is low, and no transmission congestion occurs; hence, all four DCs face nearly identical LMPs, as shown in Fig.~\ref{LMP without P2P-CSM}. In this situation, price signals alone do not motivate cloud service trading. However, the DCs exhibit heterogeneous cooling efficiencies: DC2 and DC3 operate with noticeably lower COP values than DC1 and DC4, leading to higher electricity consumption for processing the same workload. As a result, they voluntarily purchase cloud services from DC1 and DC4 through the P2P-CSM and shift part of their IWs to these more energy-efficient peers. This behavior is clearly reflected in Fig.~\ref{Workload profiles of 4 DCs with P2P-CSM}, where positive trading flows from DC2/DC3 to DC1/DC4 can be observed in the corresponding time intervals.

\begin{table}[t]

    \caption{Profit Comparison with/without P2P-CSM.}
    \begin{tabular}{cccc}
\hline
\hline
        {Profit (\$)} & {Without P2P-CSM} & {With P2P-CSM} & {Profit increase} \\
\hline
        {DC1}   & 77750  & 90623  & 16.6\% \\
        {DC2}   & 24503 & 32435 & 32.4\% \\
        {DC3}   & 16508  & 28165  & 70.6\% \\
        {DC4}   & 55564 & 62904 & 13.2\%\\
\hline
        {Total} & 174325 & 214127 &  22.8\% \\
\hline
\hline
    \end{tabular}
    \label{profit1}
\end{table}
\begin{figure}[t]
\centering
\vspace{-0.4cm}
\includegraphics[scale=0.8]{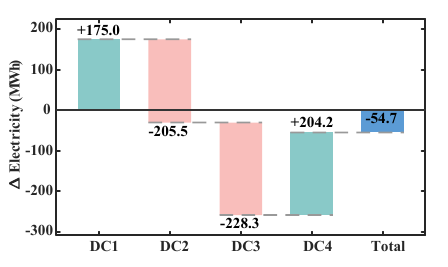}
\vspace{-0.2cm}
\caption{Electricity Consumption Variations with P2P-CSM.}
\vspace{-0.6cm}
\label{Total power}
\end{figure}
\subsubsection{Evaluation of Computation-Electricity Coordination}
This subsection evaluates the efficacy of the P2P-CSM in facilitating computation-electricity coordination and analyzes its impacts from both the DC and power grid perspectives.

From the DC perspective, Table~\ref{profit1} compares the operational profits of four DCs with and without P2P-CSM. It is observed that all DCs achieve higher profits ubder the P2P-CSM, indicating that the P2P-CSM effectively enhances the economic benefits of individual DCs while simultaneously increasing total social welfare by 22.8\%. The most significant improvement is observed at DC3, primarily because the P2P-CSM enables it to hedge against high LMPs by migrating IWs to peers with lower LMPs.

From the power grid perspective, Fig.~\ref{LMP without P2P-CSM} and Fig.~\ref{LMP with P2P-CSM} compare the LMP profiles among the four DCs. With the P2P-CSM, the local sharp price spikes at DC2 and DC3 are effectively eliminated, and all nodes exhibit a much more uniform price trajectory during the peak periods. Guided by LMP signals, the P2P-CSM facilitates workload migration toward less congested nodes, hence effectively relieves transmission network congestion and enhances the operational security and stability of the power system.
Fig.\ref{Total power} further quantifies the electricity consumption variations across DCs with P2P-CSM. Although the total volume of IWs remains unchanged, the market reshapes their spatial distribution to exploit COP disparities across different DCs. Notably, the total electricity consumption decreases by 54.7~MWh, representing a 3.2\% reduction in total electricity demand. These results confirm that the P2P-CSM effectively optimizes global energy consumption and enhances system energy efficiency.

\section{Conclusion}
\label{Conclusion}
DCs are playing an increasingly pivotal role in modern power systems. To facilitate effective coordination between DCs and the power grid, this paper first proposes a novel P2P-CSM for geo-distributed DCs, which enables bilateral cloud service transactions without a central coordinator. DCs can migrate workloads to cost-effective peers via cloud service trading, thereby enhancing their operational profit.
Second, an LMP-embedded bi-level computation-electricity coordination framework is proposed to explicitly capture the bidirectional interactions between the power grid and DCs.
Third, a dual consensus ADMM-based decentralized algorithm is developed as the P2P market clearing algorithm, and a bisection-assisted iterative algorithm is developed to address the oscillation issues and ensure rigorous convergence of the interaction.
Finally, case studies conducted on a modified IEEE 30-bus system demonstrate that the proposed P2P-CSM not only increases the total DC operational profit by 23.8\%, but also effectively relieves network congestion and reduces the total electricity consumption by 3.2\%, achieving effective computation-electricity coordination. 
Future research will extend this framework to real-time decision-making for DC workload scheduling.

\bibliographystyle{IEEEtran}
\bibliography{datacenter}


 





\end{document}